# Development of Diagnostics for High-Temperature High-Pressure Liquid Pb-16Li Applications

A. Saraswat[1], S. Sahu[1], T. S. Rao[1], A. Prajapati[1], S. Verma[1], S. Gupta[1], M. Kumar[1], R. P. Bhattacharyay[1], P. Das[2]

[1]Institute for Plasma Research, Bhat, Gandhinagar – 382 428, India
[2]Bhabha Atomic Research Centre, Trombay, Mumbai – 400 085, India

**Abstract –** Liquid lead-lithium (Pb-16Li) is of primary interest as one of the candidate materials for tritium breeder, neutron multiplier and coolant fluid in liquid metal blanket concepts relevant to fusion power plants. For an effective and reliable operation of such high temperature liquid metal systems, monitoring and control of critical process parameters is essential. However, limited operational experience coupled with high temperature operating conditions and corrosive nature of Pb-16Li severely limits application of commercially available diagnostic tools. This paper illustrates indigenous calibration test facility designs and experimental methods used to develop non-contact configuration level diagnostics using pulse radar level sensor, wetted configuration pressure diagnostics using diaphragm seal type pressure sensor and bulk temperature diagnostics with temperature profiling for high temperature, high pressure liquid Pb and Pb-16Li applications. Calibration check of these sensors was performed using analytical methods, at temperature between 380°C-400°C and pressure upto 1 MPa (g). Reliability and performance validation were achieved through long duration testing of sensors in liquid Pb and liquid Pb-16Li environment for over 1000 hour. Estimated deviation for radar level sensor lies within [-3.36 mm, +13.64 mm] and the estimated error for pressure sensor lies within 1.1% of calibrated span over the entire test duration. Results obtained and critical observations from these tests are presented in this paper.

*Keywords* - **Sensor, Lead-lithium, Liquid metal, Fusion blanket**

## 1. Introduction

One of the key missions of International Thermonuclear Experimental Reactor (ITER) is to validate the design concepts for tritium breeding blankets relevant to a power producing reactor like DEMO. ITER operation will demonstrate feasibility of breeding blanket concepts that would lead to tritium self-sufficiency and high-grade heat extraction, which are necessary goals for DEMO [1-3]. India is working towards development of Lead Lithium Ceramic Breeder (LLCB) Test Blanket Module (TBM), planned to be tested in equatorial port #2 in ITER [1]. LLCB blanket concept consists of lithium titanate ceramic breeder in the form of packed pebble bed with liquid Pb-16Li alloy eutectic (hereafter referred to as Pb-Li) acting as a tritium breeder, neutron multiplier and coolant. To achieve the intended operation of LLCB blanket, Pb-Li is circulated in a closed loop, called Lead Lithium Cooling System (LLCS), which extracts the volumetric heat generated within TBM internal structures along with its self-generated neutronic heat. LLCS is being designed to operate at a gauge pressure of 1.2 MPa with temperatures between 300°C–460°C at TBM inlet and outlet, respectively, to extract a total heat load of 0.38 MW. Process parameters like pressure, level, temperature and flow need to be monitored and controlled for an effective functioning of LLCS as a primary cooling system. However, limited instrumentation availability for liquid metal applications and aggressive operating parameters of LLCS coupled with corrosive nature of Pb-Li presents additional major challenges for process sensors and diagnostic tools. Additionally, typical interval of 2-years between scheduled maintenance cycles of ITER [4] demands precise validation and reliable performance of sensors over long duration operation. As a first step to address some of these challenges, experimental studies were carried out towards calibration and rigorous performance validation of sensors as part of process instrumentation development for static liquid Pb/Pb-Li applications. To fulfil the test objectives, calibration test facilities were designed and fabricated at Institute for Plasma Research (IPR). Calibration tests and long duration tests for over 1000 hour were carried out to assess the performance as well as operational reliabilities of pressure sensors, level sensor and temperature sensor assembly in liquid Pb/Pb-Li environments relevant to LLCS.

## 2. Sensor Selection

Application of Pb-Li is primarily confined to fusion specific studies [5]. In addition, corrosive nature and high operational temperature requirements have largely impeded development of process instrumentation for Pb-Li. Proper selection of measurement technique, sizing of sensor, engineering modification/customization of Commercial Off-The-Shelf (COTS) sensors as applicable for specific requirements, installation considerations demanded by liquid metal applications and rigorous experimental validation in Pb-Li environment should be addressed to establish compatibility, to validate application feasibility and to efficaciously develop any potential measurement technology relevant to liquid Pb-Li application.

## 2.1. Pressure Measurement

For a wetted pressure sensor in high temperature environment, the sensing element along with signal conditioning electronics must be isolated from extreme temperature [6]. To achieve this, an indirect liquid metal pressure measurement approach was developed using gas pressure measurements [7]. However, such an approach essentially requires tight control over temperature to minimise errors from temperature dependent gas pressurization in a fixed volume, which may be sometimes difficult to achieve during steady state operation. In the present study, we have adopted piezo-resistive principle based remote diaphragm seal type pressure sensors with remotely mounted electronics for pressure measurement of liquid Pb/Pb-Li. Wetted diaphragm, acting as a process isolation seal, transmits pressure upto remotely mounted sensing diaphragm of piezo-resistive sensor through a fine capillary filled with an incompressible, high temperature compatible fluid. Minimum volume displacement of capillary fill fluid ensures better dynamic response towards fast changes in pressure, which is satisfactory for LLCS operational point of view. In this study, two types of pressure sensors with 0-2 MPa (g) calibrated range were selected: a customized remote diaphragm seal type pressure sensor with a 1-inch SS-316 wetted diaphragm with 0.25 mm bore rigid stem configuration capillary containing sodium-potassium alloy (NaK) as shown in *Fig.1(a)* and remote diaphragm seal type pressure sensor with a 2-inch SS-316L wetted diaphragm with 1 mm bore flexible capillary containing high temperature compatible silicone oil as shown in *Fig.1(b)*. Both the sensors are equipped with current transmitters to provide a linear output signal in the calibrated range.

## 2.2. Level Measurement

A variety of validated level diagnostics like conductivity probes, induction sensors etc. are available for a few liquid metal applications like Na/Na-K etc. [6]. However, in Pb-Li environment, a wetted configuration level diagnostics maybe rendered ineffective over long durations due to deposition of metallic oxides, corrosion of sensor probe and bending stresses exerted by high density liquid Pb/Pb-Li.

Hence, to continuously measure level in a non-contact manner, a pulse radar level sensor equipped with a linear current transmitter was selected. It is based on the principle of measuring transit time between the emitted microwave pulse and its reflected echo from the sensed material surface. The transit time is directly proportional to distance between sensor and the medium [8]. Unlike ultrasonic level sensors based on time of flight principle, radar level sensor provides a non-contact technique virtually unaffected by variations in process temperature, pressure or cover gas composition [8]. To achieve a highly focused beam and smaller process connection size, basic sensor working on 26 GHz (K-band) frequency with a pulse repetition frequency of 3.6 MHz was selected. Isolation of electronics from high process temperature was achieved by using a temperature isolator section. Antenna impedance cone of ceramic with graphite seal was selected for high temperature application. Selected sensor configuration is shown in *Fig.1(c)*.

## 2.3. Bulk Temperature Measurement and Temperature Profiling

To measure bulk temperature of liquid Pb-Li and to study feasibility of Pb-Li level estimation using vertical bulk temperature profiling, a compact Temperature Level Probe (TeLePro) assembly was developed using a K-type multilevel thermocouple as shown in *Fig.1(d)*. TeLePro provides a compact rigid probe for vertical temperature profiling of bulk medium which is adaptable for tanks with small diameter and tanks with internal installations. For this study, 21 temperature measuring junctions, separated by 20 mm each, were accommodated in a common outer sheath of length 400 mm and outer diameter of 6 mm. Junction-1 corresponds to tip of sensor probe. A properly sized Inconel-600 thermowell protects the sensor probe from corrosion, process pressure and bending stresses. Level estimation using TeLePro is based on differential temperature measurement technique and can be further adapted to enhance the response and resolution by proper thermowell sizing and location of sensing points, respectively. Sensor customization might be limited by manufacturing feasibility and detectable temperature gradients.

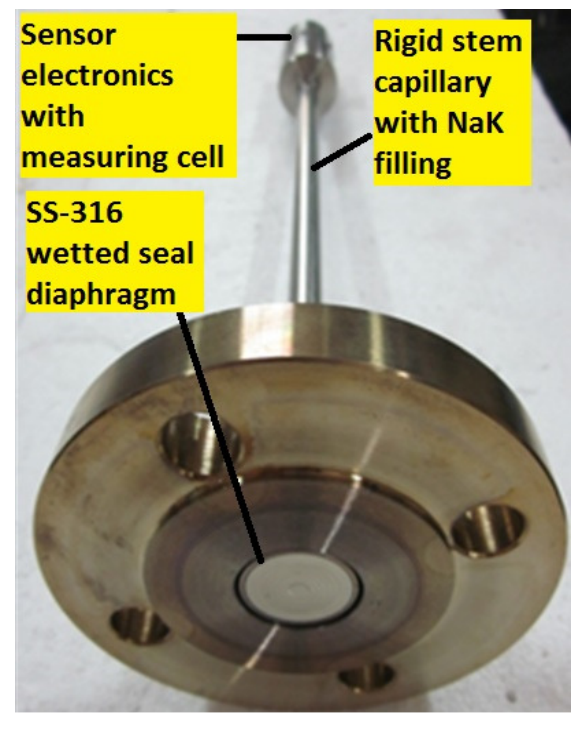

*Fig.1(a): NaK fill fluid based remote diaphragm seal type pressure sensor.*

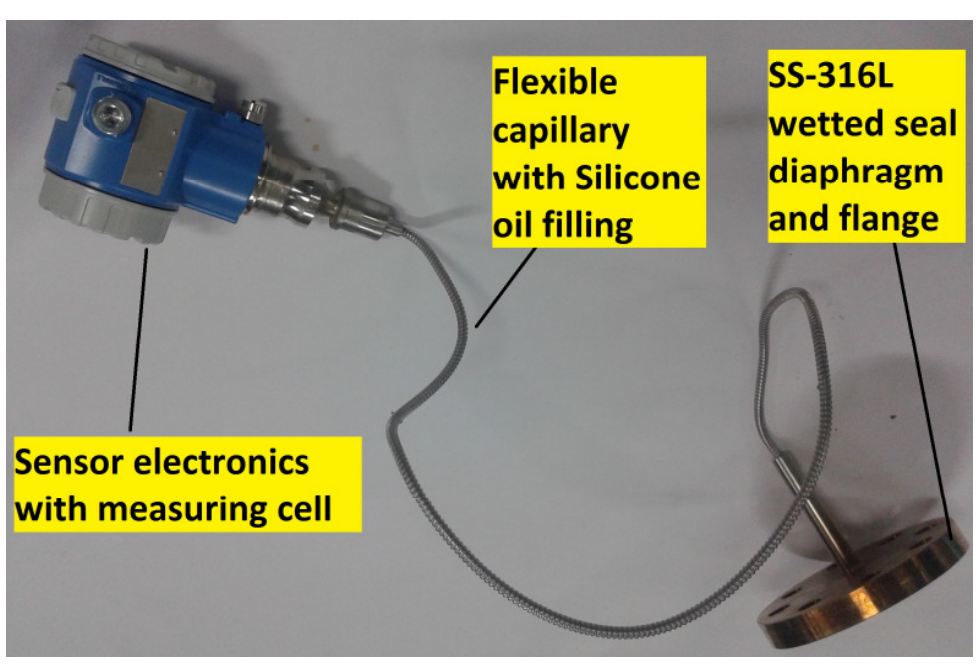

*Fig.1(b): High Temperature compatible silicone oil fill fluid based remote diaphragm seal type pressure sensor.*

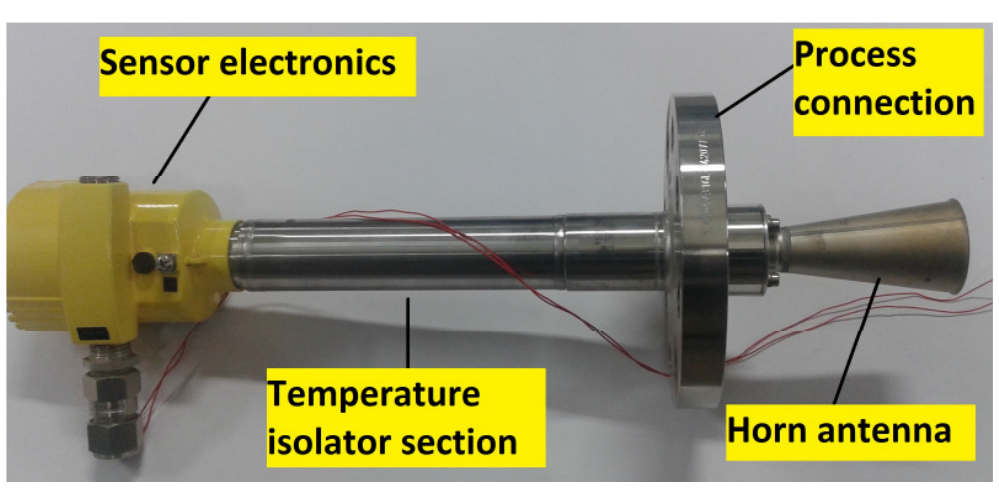

*Fig.1(c): Non-contact type pulse radar level sensor with temperature isolator section to decouple high process temperatures.*

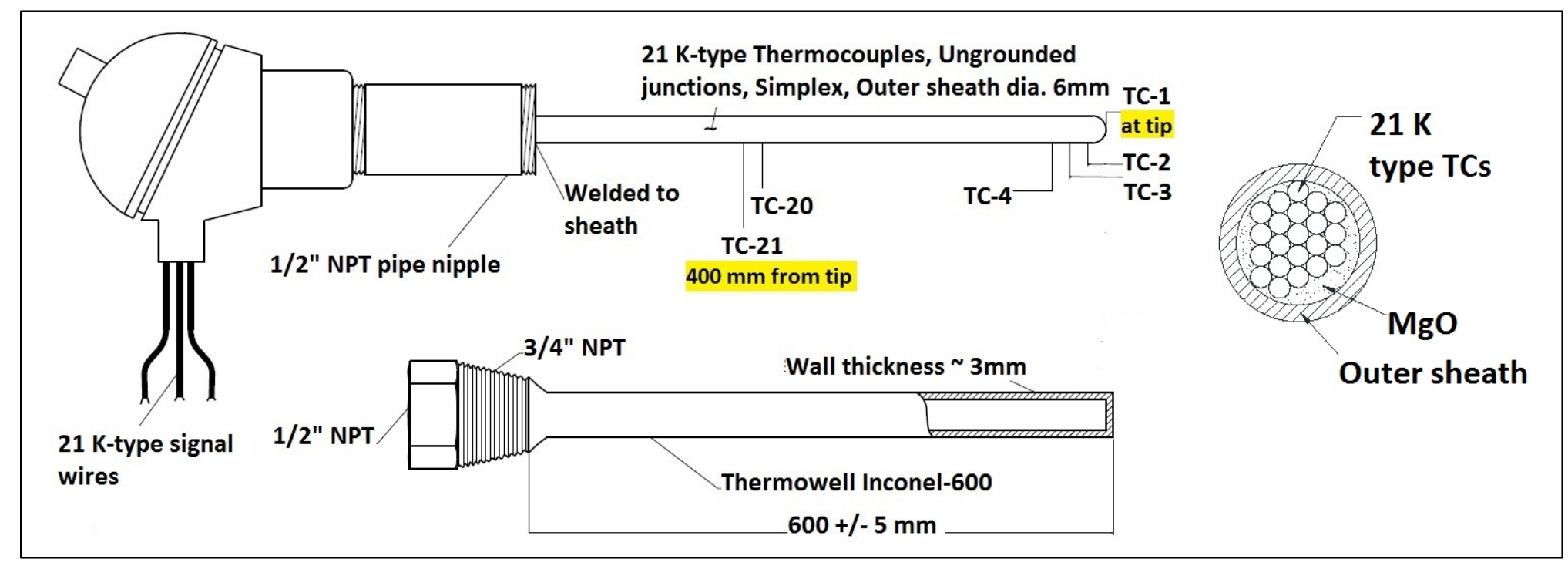

*Fig.1(d): TeLePro assembly with multiple temperature measuring junctions (K-type) for bulk temperature measurements and liquid metal level estimation.*

## 3. Experimental Details

Two calibration test facilities were in-house designed for experimental validation of selected pressure, level and temperature sensors. These test facilities were designed to provide required test environments as well as to address installation constraints of selected sensors.

### 3.1. Test Facility-1: Design Description and Test Procedure

Test facility-1 was designed for calibration and performance validation of selected radar level sensor and silicone oil filled pressure sensor for liquid metal application. Major components of test facility-1 include main tank, top nozzle, side-section, drain tank and connecting pipeline with an isolation valve. Material of construction for these components is SS-316L. Schematic diagram of test facility-1 is shown in *Fig.2(a)* and an image of process sensors installed on the main tank is shown in *Fig.2(b)*. Table-I lists major process test parameters. Main tank height (600 mm) was derived from the height of inventory tank planned for LLCS, while internal diameter of main tank (492 mm) and height of connection nozzle were primarily governed by installation constraints of level sensor to avoid false echoes. To achieve sufficient level variations in the large volume of main tank, liquid Pb was selected as an economical surrogate process medium. Drain tank, with an internal diameter of 202.7 mm and height of 630 mm, allows for variation of liquid Pb level using cover gas pressurization technique. Side-section, with an adequate inclination for gravity assisted liquid metal draining, is provided for installation of silicone oil filled pressure sensor. Controlled shifting of liquid Pb between the tanks was achieved through manual operation of isolation valve. A secondary reference for liquid metal level estimation was established using conductivity level switches for discrete level measurements (*Fig.3*). Porcelain insulator of spark plug provided electrical isolation of electrode from tank structure.

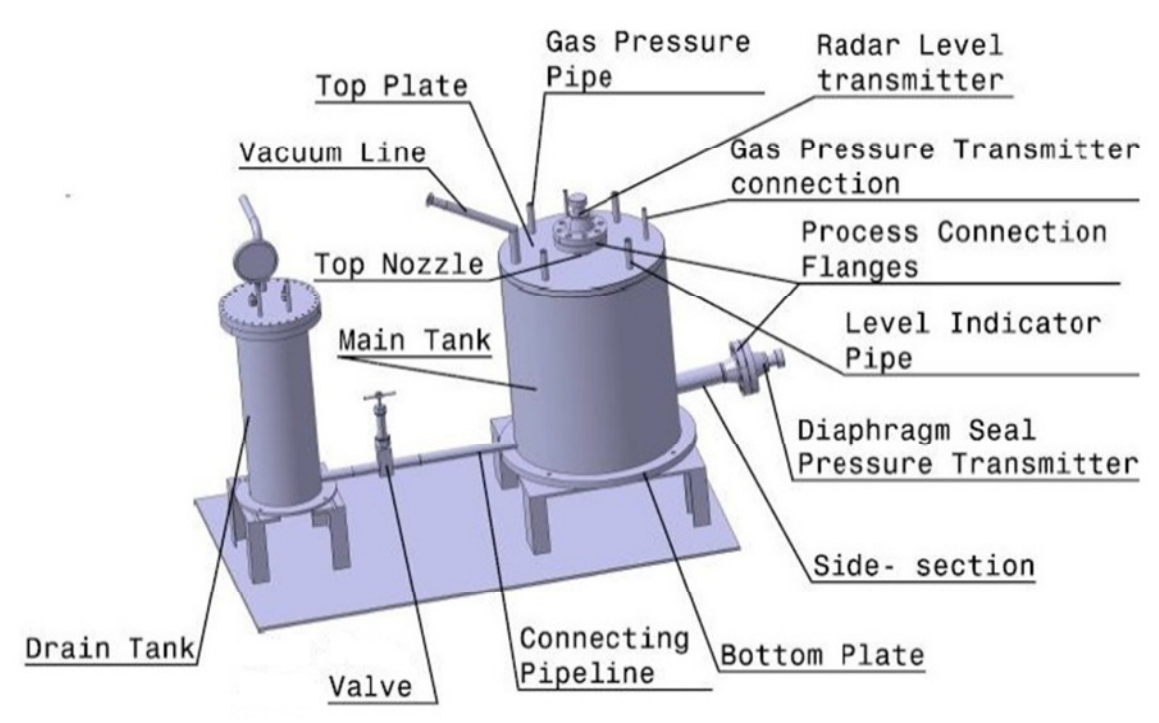

*Fig.2(a): Schematic diagram for test facility-1.*

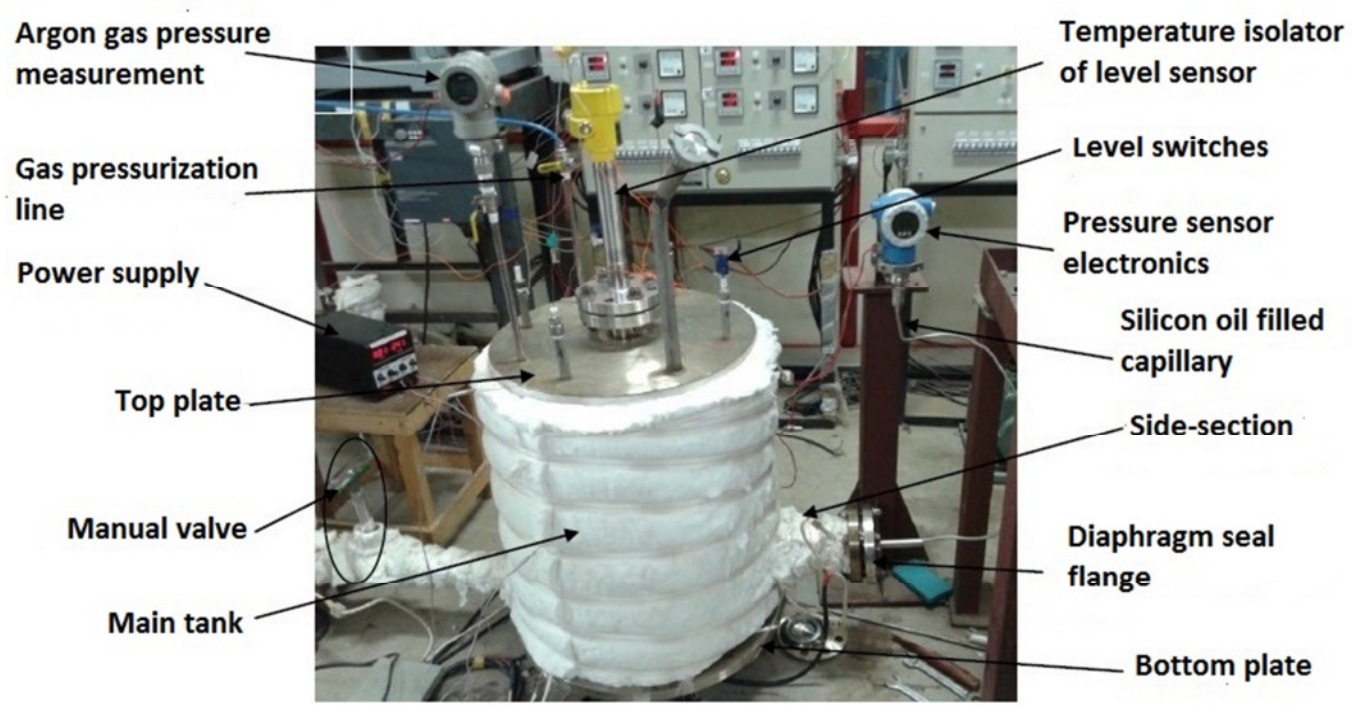

*Fig.2(b): Image of test facility-1 main tank. Level sensor and pressure sensor are installed on the top nozzle and side-section, respectively.*

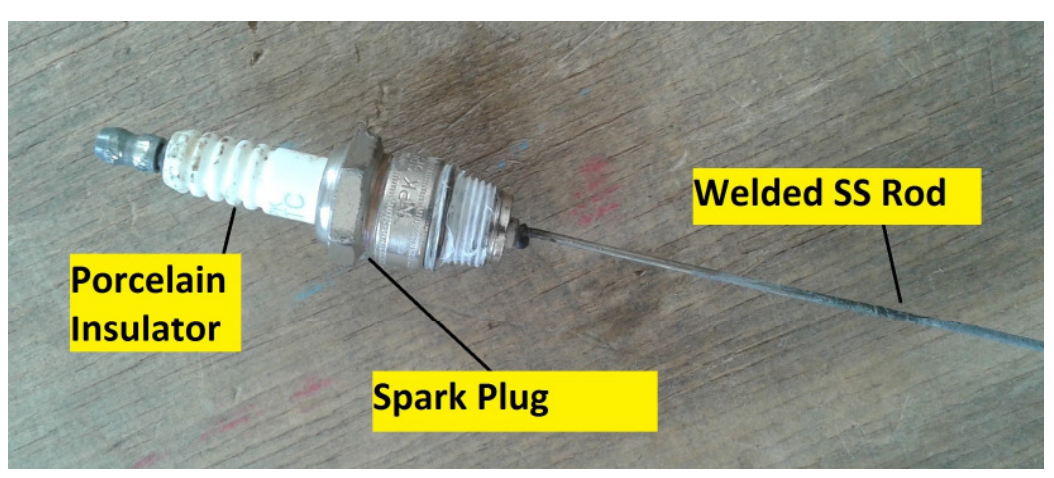

*Fig.3: Conductivity level switch construction.*

TABLE-I: PROCESS PARAMETERS FOR TEST FACILITY-1

| Process Medium | Liquid Pb |
|---|---|
| Operating Temperature | 380 ºC - 400 ºC |
| Operating Pressure | Upto 1 MPa (gauge) |
| Density of Pb at 400ºC | 10,584 kg/m$^3$ [9] |
| Melting point of Pb | 327.4 ºC [9] |

For validation on liquid Pb application, radar level sensor was configured to measure level ($H$) in the range of 0-694 mm as per the height of main tank with top nozzle. A known inventory (405 kg) in the form of solid Pb ingots was melted in main tank in an inert environment by applying positive pressure of argon cover gas. A reference calibrated gauge pressure sensor mounted on main tank measured cover gas pressure ($P_g$). To perform first point calibration check, level of liquid Pb in main tank was analytically estimated using mass of Pb taken, density of Pb at operating temperature and dimensions of test facility-1 and compared with output of radar level sensor. For second point calibration check, liquid Pb was transferred to drain tank upto a fixed level, estimated using a fixed conductivity level switch. The level of liquid Pb remaining in main tank was calculated and compared with the output of radar level sensor. All calculations take into account thermal expansion of both tanks along height to correct for zero level offset of radar level sensor and to correctly estimate liquid Pb level at operating temperature. Estimated radial expansions are negligible due to welded top and bottom plates. After calibration checks, liquid Pb inventory was transferred to main tank for long duration test. Test was performed for over 700 hour continuously with cover gas pressure upto 1 MPa (g) and readings were taken at regular intervals to validate the performance of radar level sensor.

For pressure sensor testing, diaphragm seal was installed on the side-section port. Zero adjustment for the elevation of sensor electronics mounting position was carried out. Sensor diaphragm seal was located 145 mm above the bottom of main tank, hence only a part of liquid Pb column ($H_{effective}$) generated pressure ($P_{effective}$) on the pressure sensor diaphragm.

$$H_{effective} = H - 145 \text{ mm} \quad \ldots\ldots\ldots\ldots\ldots\ldots \quad (i)$$

$$P_{effective} = H_{effective} \cdot \rho_{medium} \cdot g \quad \ldots\ldots\ldots\ldots\ldots\ldots \quad (ii)$$

where $\rho_{medium}$ is density of Pb at operating temperature and $g$ is acceleration due to gravity. Total pressure ($P$) exerted on diaphragm seal is summation of pressure applied through cover gas ($P_g$) and pressure head due to effective liquid Pb column ($P_{effective}$).

$$P = P_{effective} + P_g \quad \ldots\ldots\ldots\ldots\ldots\ldots \quad (iii)$$

From equation *(iii)*, we can see that by varying cover gas pressure alone, one can vary total pressure applied to pressure sensor under test, while simultaneously ensuring diaphragm seal in direct contact with liquid Pb. Calculated total pressure $P$ was compared with pressure sensor output to estimate the error. Testing was continuously done for over 310 hour and readings were taken at regular intervals.

### 3.2. Test Facility-2: Design Description and Test Procedure

Test facility-2 was fabricated for experimental validation of pressure sensors and TeLePro assembly in high-temperature, high-pressure liquid Pb-Li environment. This study was intended to test compatibility of sensors with liquid Pb-Li, to understand deteriorating effects of corrosion on measurement reliabilities and to check feasibility of liquid Pb-Li level estimation using TeLePro. Tank-A, with an internal diameter of 77.9 mm and height of 412 mm, was used to test both types of pressure sensors simultaneously on two side-sections while tank-B, with an internal diameter of 52.5 mm and height of 528 mm, was used for testing of TeLePro assembly. Tubes were introduced between end of each side-section and tank-A top volume to remove trapped gas volume during charging of Pb-Li as well as to help ensure proper drainage of liquid Pb-Li from side-sections using active gas pressurization. Testing of pressure sensors and TeLePro assembly were performed sequentially as shown in *Fig.4(a)* and *Fig.4(b)*. Process test parameters are listed in Table-II.

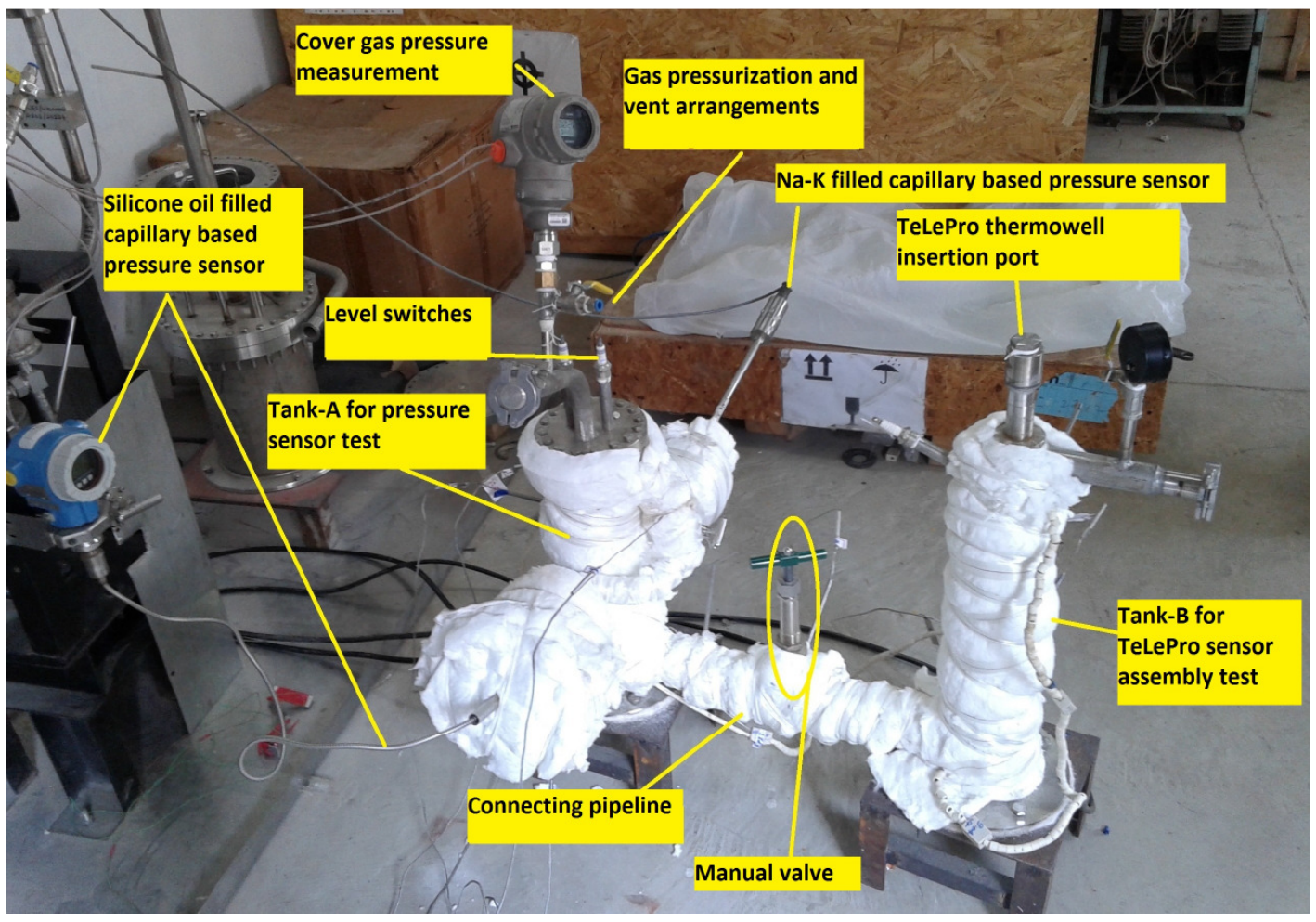

*Fig.4(a): Pressure sensors testing phase in test facility-2.*

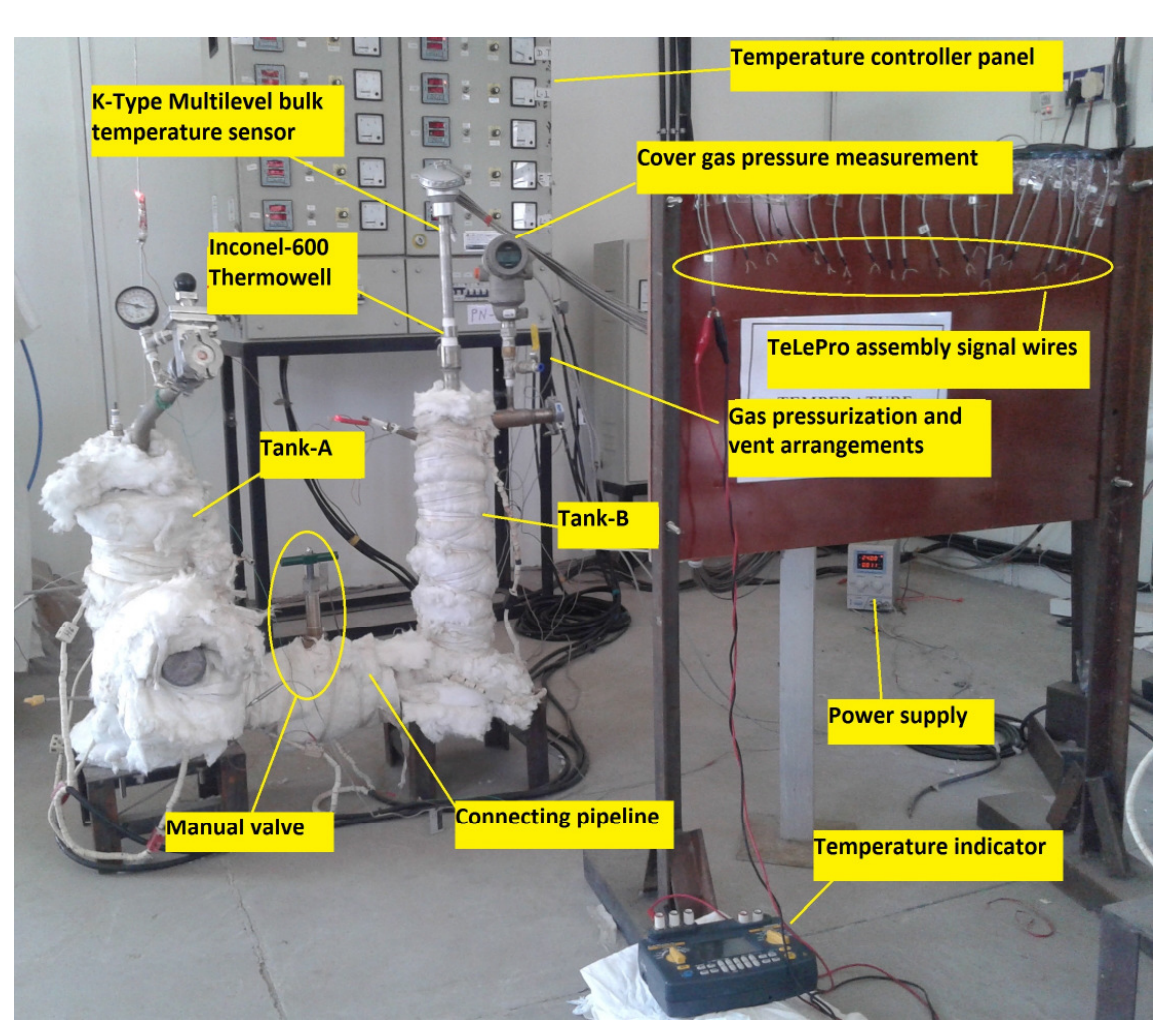

*Fig.4(b): TeLePro assembly testing phase in test facility-2.*

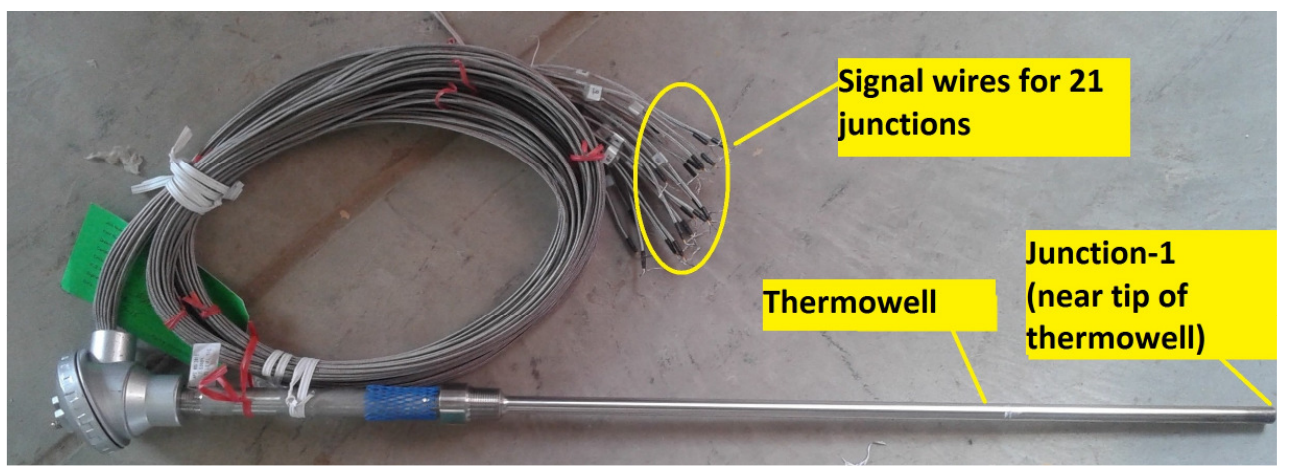

*Fig.5: TeLePro assembly before exposure to liquid metal.*

TABLE-II: PROCESS PARAMETERS FOR TEST FACILITY-2

| Process Medium | Liquid Pb-16Li |
|---|---|
| Operating Temperature | 250 ºC - 530 ºC |
| Operating Pressure | Upto 1.05 MPa (gauge) |
| Density of Pb-16Li at 400ºC | 9318 $kg/m^3$ [10] |
| Melting point of Pb-16Li | 235 ºC [10] |

For pressure sensors calibration and testing, temperatures of both the side-sections were maintained between 380ºC - 400ºC. Effective Pb-Li head for each pressure sensor was estimated through an average of repeated measurements done using movable conductivity type level switch. For silicone oil filled pressure sensor, $H1_{effective}$ was 66 mm and for NaK filled pressure sensor, $H2_{effective}$ was 75 mm. Total exerted pressures on the pressure sensors were estimated using equations *(ii)* and *(iii)* where density of Pb-Li at 400ºC was estimated using empirical relation [10]. Calculated total exerted pressures were compared with sensor outputs to estimate the error. Two calibration cycles, each from 0 to 1 MPa (g) and vice-versa, were performed at start and end of continuous 1000 hour long duration test. Cover gas pressure was maintained in tank-A and readings were taken at regular intervals.

The customized TeLePro assembly is shown in *Fig.5*. TeLePro was subjected to severe thermal aging in corrosive liquid Pb-Li with cover gas pressure upto 1 MPa (g). For first 500 hour, heater of tank-B was controlled using surface temperature measurement and for next 500 hour, junction-1 of TeLePro was used for heater control using bulk Pb-Li temperature measurement. Maximum bulk temperature upto 520ºC was attained for extended durations. After completion of continuous 1000 hour long duration test, TeLePro development campaign for liquid Pb-Li level estimation was carried out. In principle, a reliable level sensing technique should be unaffected by process temperature and pressure conditions. To validate the assembly, steady state temperature profiles were obtained for two cases: *(i)* different cover gas pressures at a constant temperature Control Set-Point (CSP) and *(ii)* different temperature CSPs at a constant cover gas pressure. Sensor assembly was exposed to liquid Pb-Li continuously for over 1240 hour. After completion of testing, TeLePro assembly was removed in hot condition from the facility.

## 4. Results and Discussions

For first point analytical calibration of radar level sensor, estimated liquid Pb level in main tank was 198.42 mm and observed sensor reading was 204.97 mm. Thus a deviation of +6.55 mm was observed. For second point calibration, analytically estimated level in main tank was 104.28 mm and observed sensor reading was 116.66 mm. Hence, a deviation of +12.38 mm was observed. It should be noted that before employing radar level sensor on liquid Pb application, calibration checks were performed under ambient conditions using *(i)* a metallic reflector plate and *(ii)* water as process fluid. During these tests, observed error was within [+1 mm, +5 mm], which is close to as suggested by manufacturer's datasheet. A few echo curves were also obtained during liquid Pb level measurements. *Fig.6(a)* shows an observed echo pattern from liquid Pb surface alongwith preset false signal suppression curve.

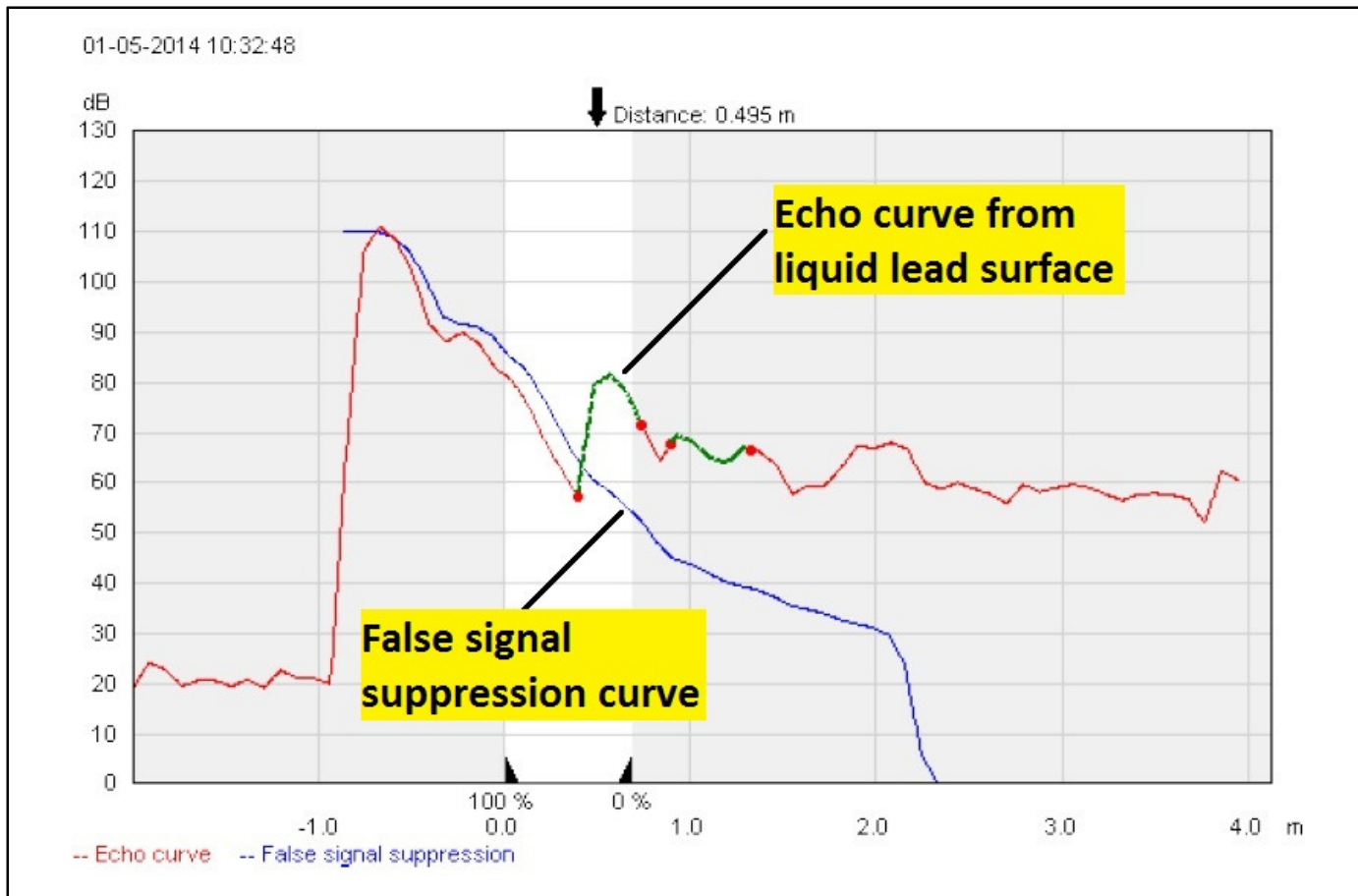

*Fig.6(a): Echo observed at 495 mm (from radar level sensor reference plane) during liquid Pb level measurements (preset false signal suppression curve is also shown).*

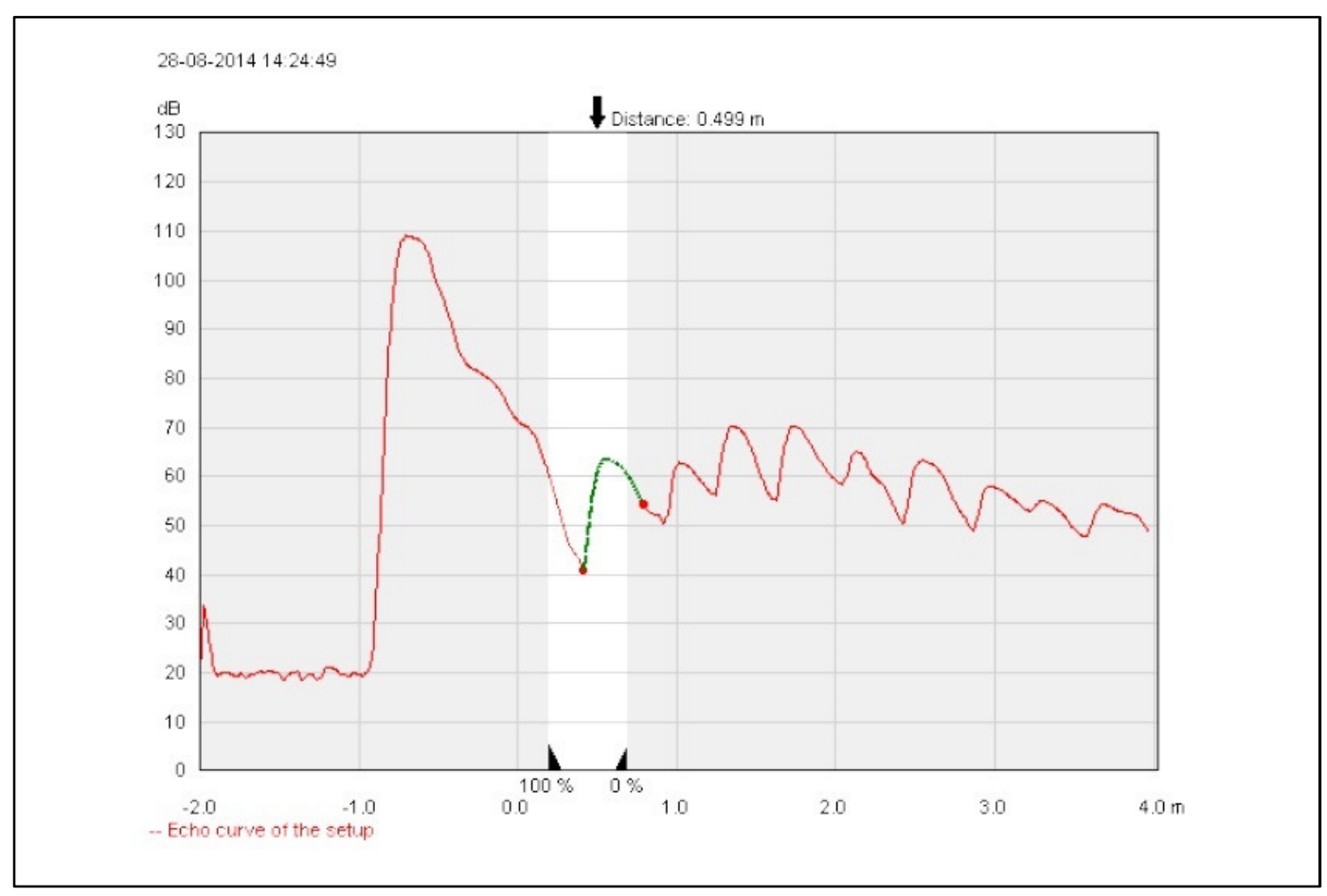

*Fig.6(b): Echo observed at 499 mm (from radar level sensor reference plane) during liquid Pb level measurements after removing false signal suppression curve.*

This echo suggests an average distance of liquid Pb surface at 495 mm from reference plane of top nozzle, which corresponds to liquid Pb level of ~203 mm from main tank bottom. During design, all sources of false echoes were taken care of, hence preset false signal suppression curve was removed to obtain clean liquid metal echo of interest as shown in *Fig.6(b)*, which suggests liquid Pb level of ~199 mm. In both the cases, observed echo amplitudes (> 60 dB) affirm strong reflected echoes. *Fig.7* presents long duration test data for liquid Pb level measurement. The obtained data suggests absence of smooth surface which may be attributed to level readings averaged over a surface area containing floating oxide layers with multifaceted surface topography on top of the melt. Estimated deviation in level measurement of liquid Pb was within [-3.36 mm, +13.64 mm] for over 1000 hour duration. Sources of error include assumption of constant bulk density of liquid Pb for analytical estimation of true level, error in level detection using conductivity switch and measurement accuracy of sensor itself. However, accuracy of radar level sensor re-confirmed on water application was better than 3 mm, while errors due to other sources were estimated very small. Hence, observed deviation seems largely due to irregular surface profile of oxide layers. Presence of such irregular oxide layers was later visually confirmed in similar applications. In future experiments, a separate melt tank is planned to melt and transfer clean liquid metal in the main inventory tank. As the selected radar level sensor is non-contact configuration, its calibration and rigorous validation for liquid Pb also corroborates its suitability for other liquid metals/metallic alloys including Pb-Li.

Estimated error in pressure measurement of liquid Pb using silicon oil filled pressure sensor was within 0.3% of calibrated span for over 310 hour test duration. *Fig.8* presents long duration test data for pressure measurement of liquid Pb.

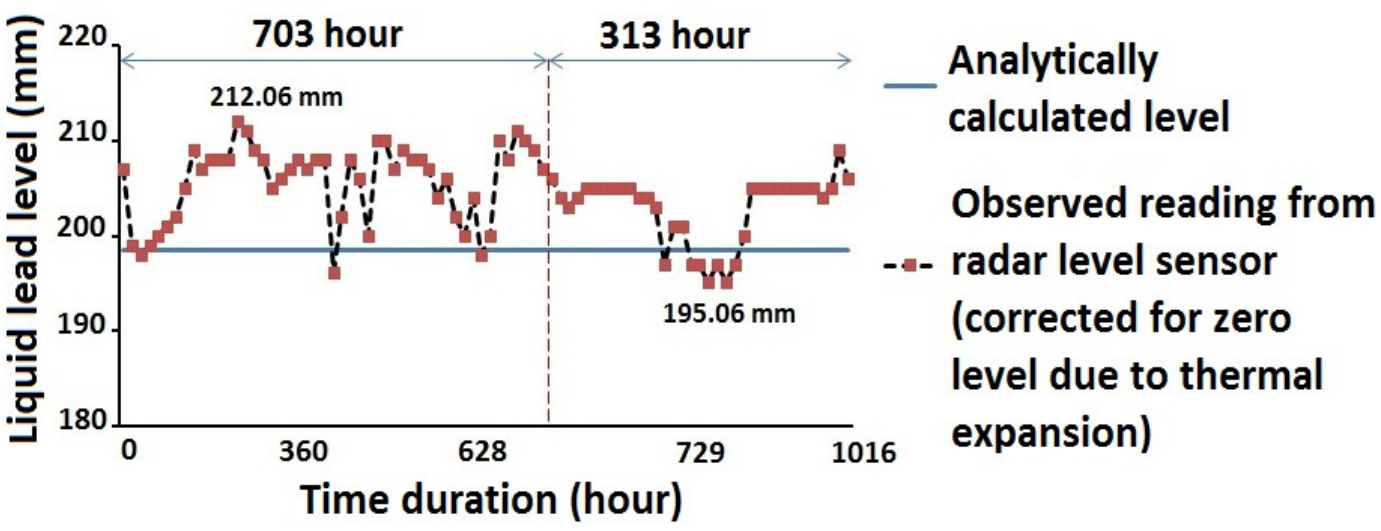

*Fig.7: Long duration test data from non-contact radar level sensor suggests absence of smooth liquid Pb surface.*

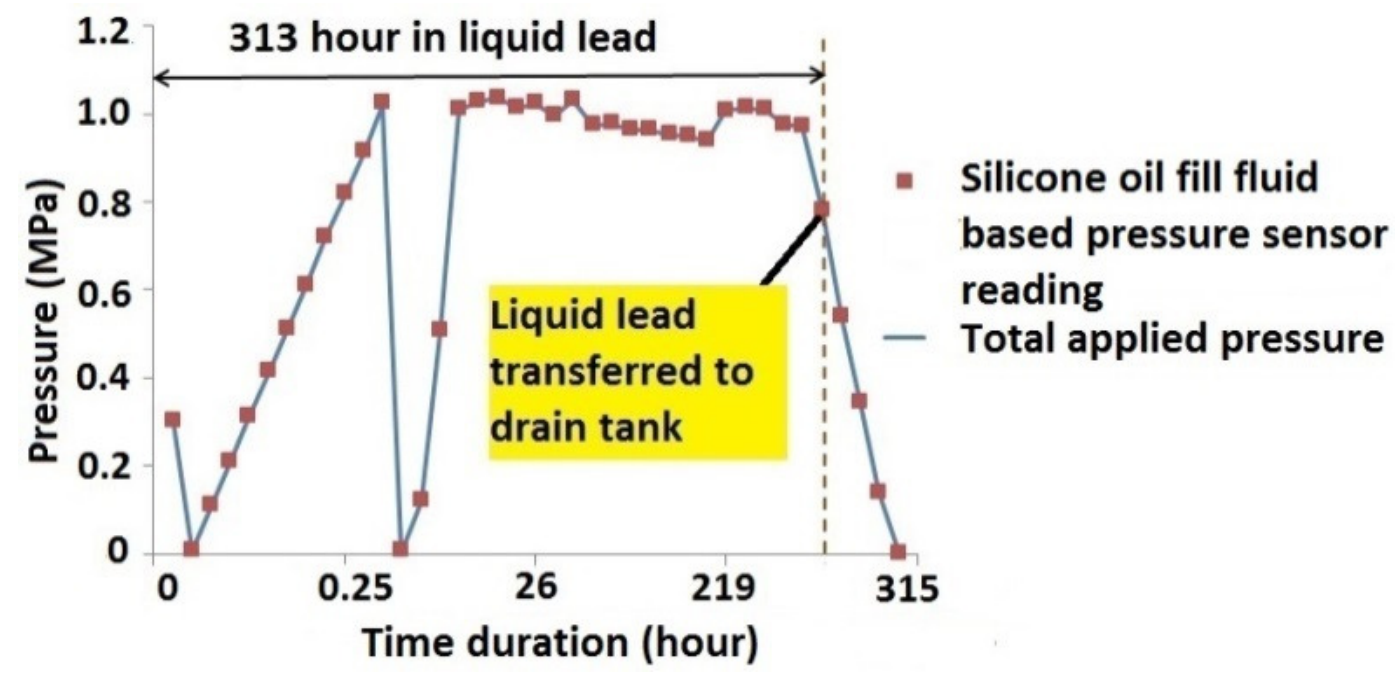

*Fig.8: Calibration cycle and long duration test data for silicone oil fill fluid based pressure sensor on liquid Pb.*

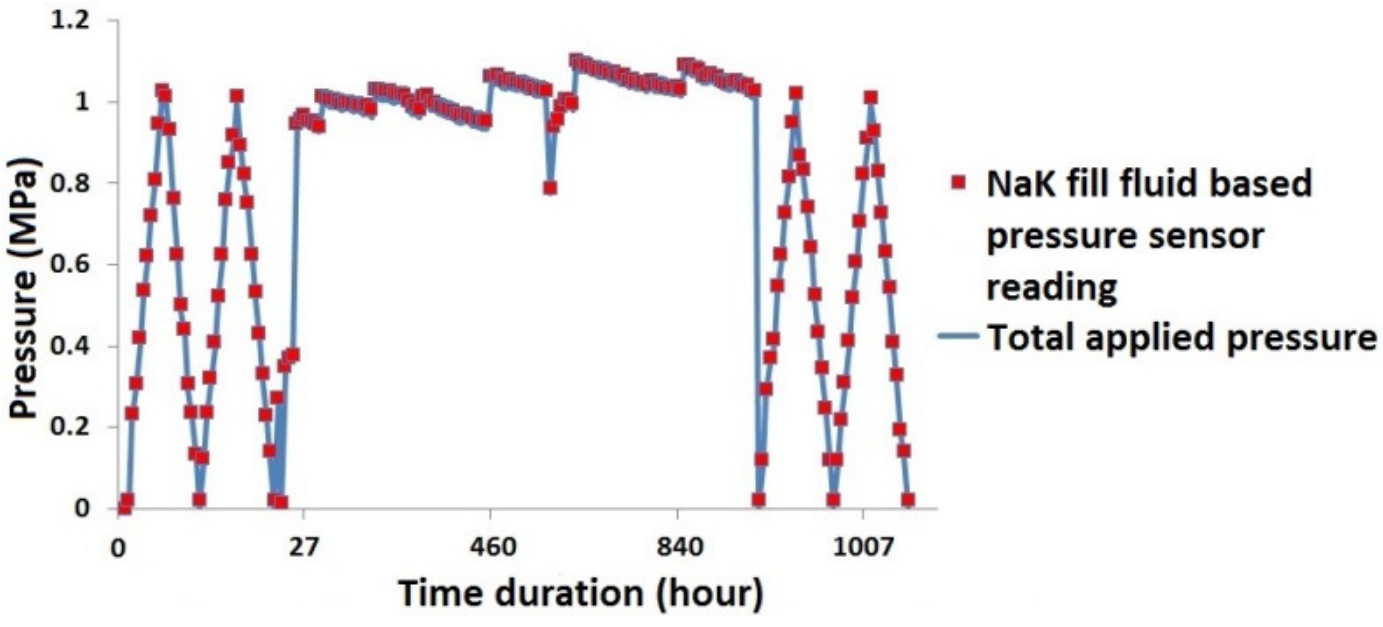

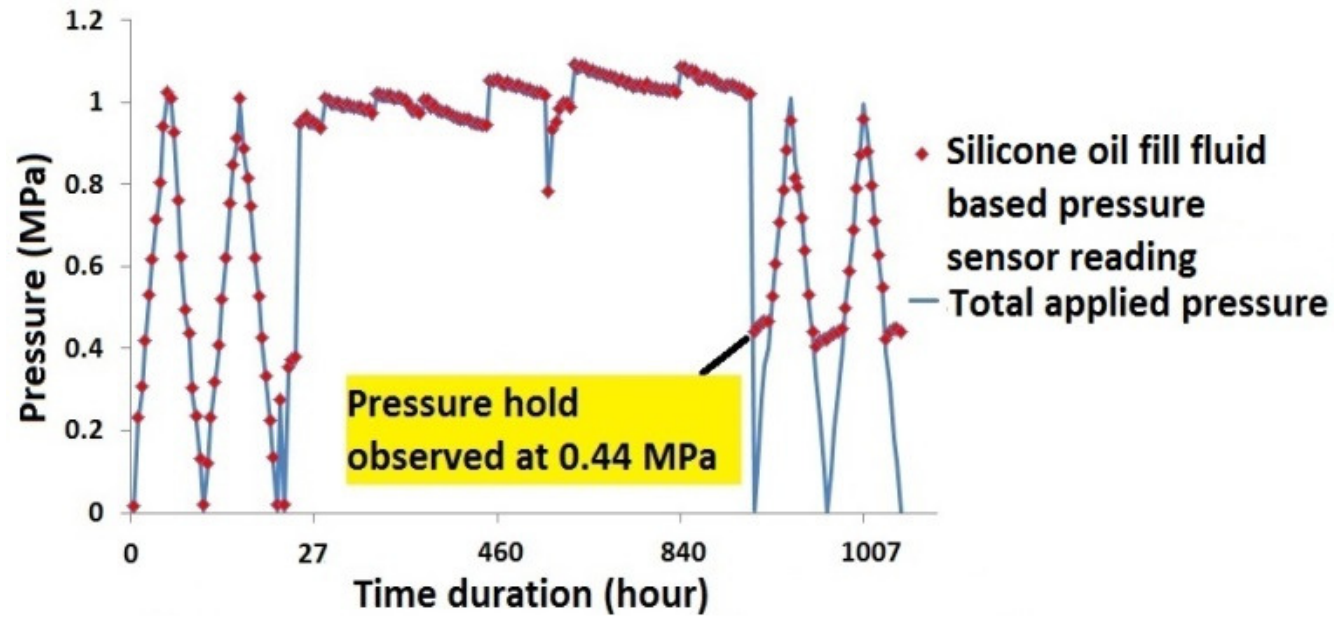

*Fig.9(a): Long duration test data for NaK fill fluid based and silicone oil fill fluid based pressure sensors on Pb-Li (gradual changes in applied pressure represent decrease of pressure with time). M-shaped cycles at the start and end of long duration test represent calibration cycles from 0 to 1 MPa (g) and vice-versa.*

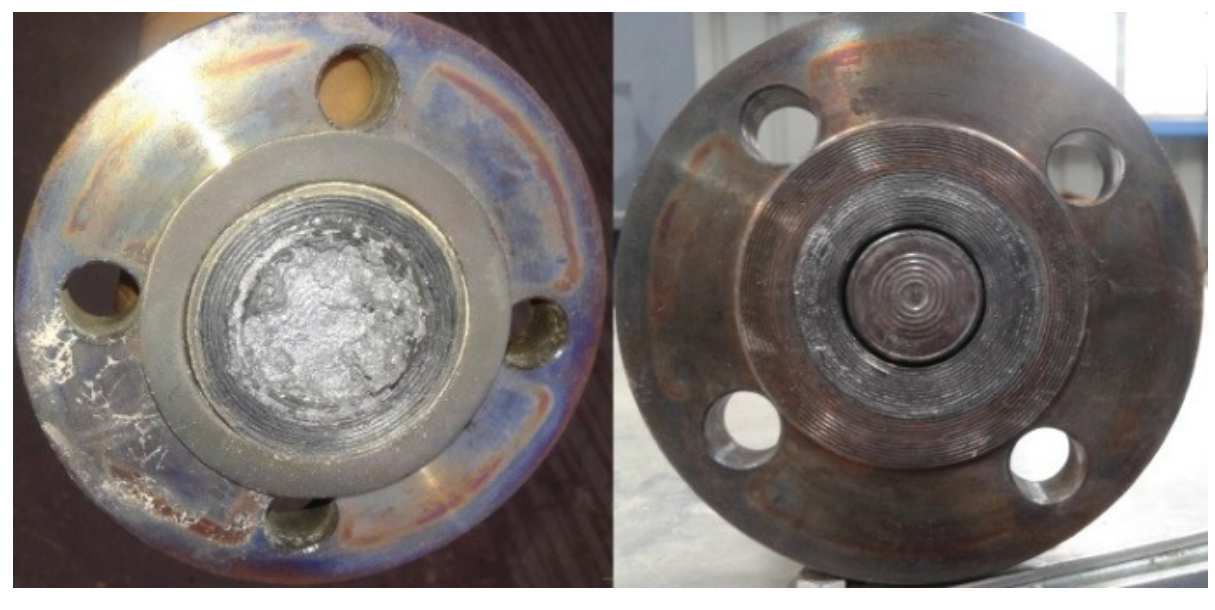

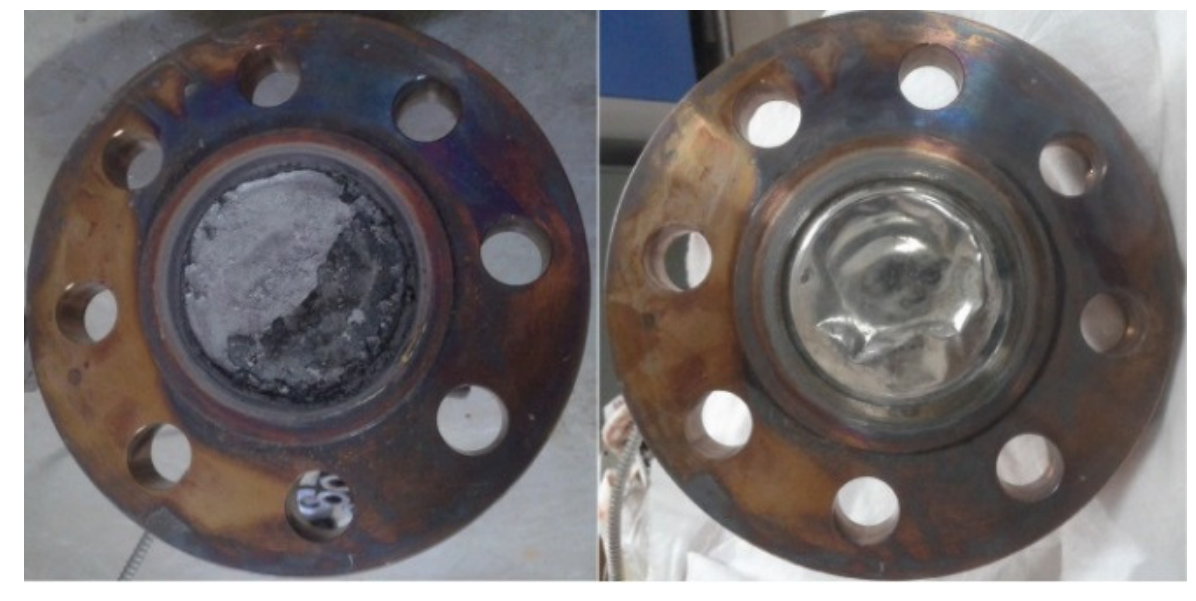

*Fig.9(b): Condition of wetted diaphragm seals for NaK fill fluid based and silicone oil fill fluid based pressure sensors after over 1000 hour continuous exposure to liquid Pb-Li and after chemical cleaning.*

*Fig.9(a)* and *Fig.9(b)* show test results obtained and condition of diaphragm seals after long term exposure to Pb-Li at high temperature and pressure. NaK filled pressure sensor showed good agreement with total applied pressure for complete test duration including calibration cycles and the estimated error lies within 1.1% of calibrated span for over 1000 hour test duration. Silicone oil filled pressure sensor showed good agreement with total applied pressure during initial two calibration cycles and during long duration test. Estimated error lies within 0.9% of calibrated span for this test duration. But at start of third calibration cycle, silicone oil filled pressure sensor did not follow total applied pressure and displayed a gauge pressure reading between 0.4 MPa - 0.46 MPa for all applied pressures less than 0.4 MPa. Visual inspection confirmed deposition of Pb-Li (with lustrous silvery characteristics) over seal diaphragm of NaK filled pressure sensor while seal diaphragm of silicone oil filled pressure sensor was partially covered with lustrous Pb-Li and partially with a greyish-blackish layer. X Ray Diffraction (XRD) analysis of samples taken from corresponding side-sections indicated presence of PbO and $Li_2O$. This oxide formation can occur due to various factors including presence of free oxygen, $Li_2O$ and PbO in the used Pb-Li ingots and surface oxidation of ingots during long term storage. The abrupt behaviour observed for silicone oil filled pressure sensor could be due to damage/distortion of seal diaphragm, formation and deposition of oxides on the seal diaphragm, thermal expansion of silicone oil inside the capillary or could be a combination of one or more of the above factors. Further diagnosis was carried out by providing heat to diaphragm seal at ambient pressure, where pressure sensor displayed a substantial increase (upto 0.38 MPa gauge) in measured value with rise in temperature at ambient pressure. This behavior suggests thermal expansion of silicone oil leading to pressurization of measuring cell membrane.

For further inspection, diaphragm seals were chemically cleaned by immersing in an equal volume mixture of acetic acid, hydrogen peroxide and ethyl alcohol followed by cleaning with water [11]. The diaphragm seal of silicone oil filled pressure sensor displayed signature of distortion. In a similar trial with an identical silicone oil filled pressure sensor, after 160 hour of continuous exposure to liquid Pb followed by continuous 210 hour in liquid Pb-Li, a similar pressure reading hold was observed between 0.42 MPa - 0.44 MPa for all applied pressures less than 0.4 MPa, suggesting prominent thermal expansion of silicone oil inside the capillary at temperatures near 400 ºC. These observations indicate ineffectiveness of silicone oil fill fluid for long duration operations in high temperature liquid Pb/Pb-Li applications relevant to LLCS.

For TeLePro development as a liquid Pb-Li level estimation technique, tank-B surface heater was controlled using a single point bulk Pb-Li temperature measurement with junction-1 of TeLePro. Higher thermal conductivity of Pb-Li would allow more heat transfer from constantly heated tank walls to the TeLePro through bulk Pb-Li. While, above Pb-Li melt, due to presence of oxides layer and cover gas region with relatively lower thermal conductivity, transferred heat towards TeLePro would be less, leading to a reduction in measured temperature by TeLePro junctions. Hence a sharp decrease in temperature was expected in oxides and cover gas region leading to interface detection. Steady state temperature profiles obtained are shown in *Fig.10(a)* and *Fig.10(b).*

Observed temperature profiles depict thermal stratification within the thermally insulated tank-B, in which static Pb-Li column is arranged into layers of different densities. Hence, an increasing temperature profile is observed within bulk Pb-Li starting from junction-1 upto the Pb-Li surface.

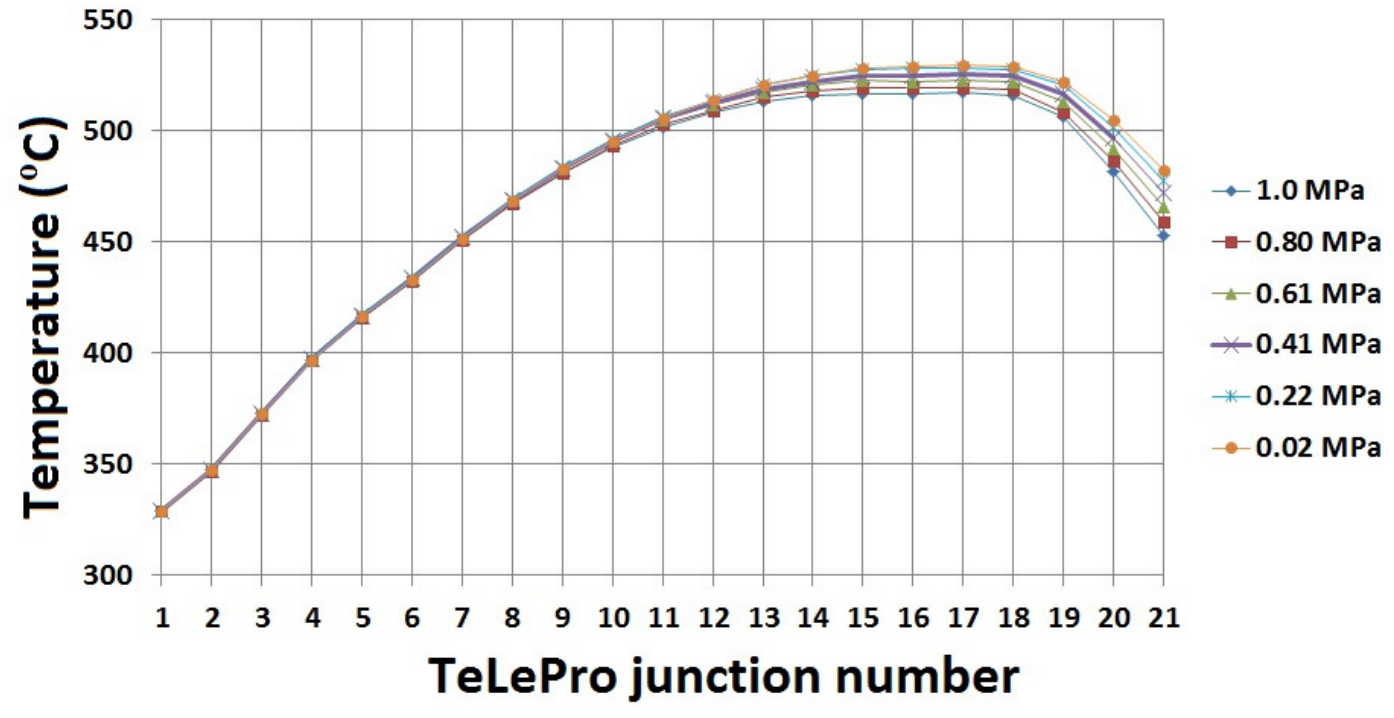

*Fig.10(a). Case-I: Temperature profiles at constant CSP of 330ºC with different cover gas pressures.*

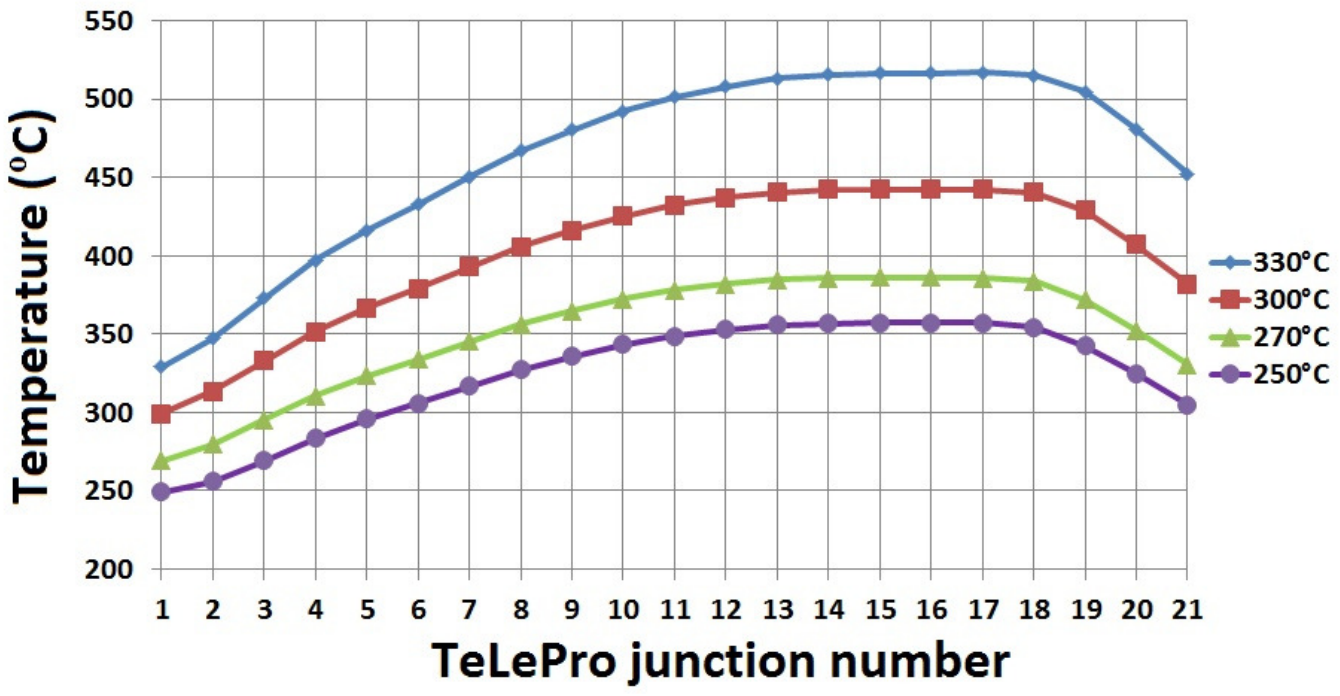

*Fig.10(b). Case-II: Temperature profiles at constant gas pressure of 1 ± 0.01 MPa (g) with different CSPs.*

As observed in Case-I, temperature for the region near Pb-Li top surface decreased with an increase in cover gas pressure while bulk Pb-Li temperature measured by lower junctions remained more or less the same. Observed profiles could be explained by the fact that the number of gas molecules increased with increase in gas pressure, aiding in more heat removal from Pb-Li surface region. In contrast, during Case-II, temperature profiles show an overall shift upwards with an increase in temperature CSP due to an effective bulk temperature rise for Pb-Li, oxides region and cover gas. Obtained profiles from both the cases show that temperature increased continuously from TeLePro junction-1 to junction-15, remained nearly constant (within 3°C) from junction-15 to junction-18 and thereafter decreased continuously. For all observed cases, obtained temperature drops were $\Delta_{18\text{-}19}$ = 7°C to 12°C, $\Delta_{19\text{-}20}$ = 17°C to 25°C and $\Delta_{20\text{-}21}$ = 20°C to 29°C where $\Delta_{a\text{-}b}$ denotes temperature drop from junction *a* to junction *b*. Observed pattern and magnitude of temperature drop indicated liquid Pb-Li surface below junction-19, but very close to junction-19. Junction-19 lies at 366.4 mm from the thermowell tip as per TeLePro design. TeLePro assembly was removed after over 1240 hour of continuous exposure and a clear demarcation was observed on the thermowell with deposited greyish layers starting at a distance of nearly 370 mm from thermowell tip. This visual observation further corroborates that liquid Pb-Li level estimation can be done within reasonable accuracy limits using TeLePro. In the proposed sensor assembly, precise level estimation is governed by resolution, which is defined by the physical separation between successive temperature sensing junctions.

*Fig.11(a)* shows observed deposited layers on TeLePro thermowell. XRD analysis indicated PbO and $Li_2O$ in the top deposited greyish layers while no evidence of nickel was present in the samples taken from different locations along the length of thermowell, suggesting negligible corrosion over this test period. Sensor probe was observed completely clean and protected from corrosion. The obtained preliminary results from TeLePro tests are encouraging. However, further tests are required for calibration of TeLePro against a reference level sensor to detect interface between bulk liquid Pb-Li and cover gas/oxide layers on the top of Pb-Li.

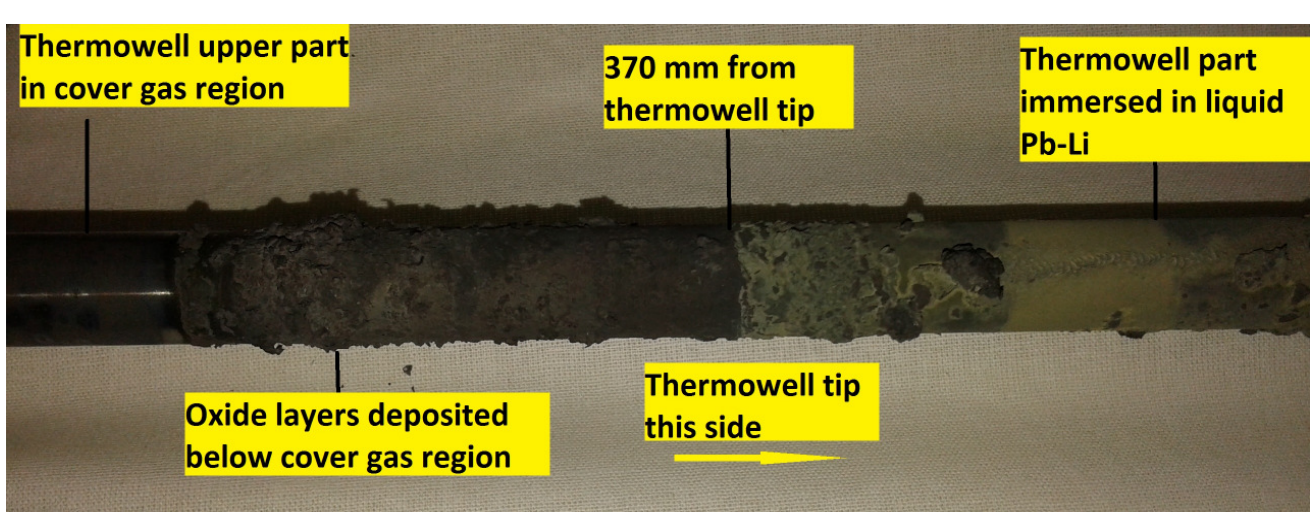

*Fig.11(a):Observed deposition of greyish oxide layers starting at 370 mm (from the tip) on TeLePro thermowell.*

## 5. Conclusion

Since instrumentation availability is limited for liquid metal applications, rigorous test methods were applied to develop and validate pressure, level and bulk temperature diagnostics for static liquid Pb/Pb-Li applications. Indigenous calibration test facilities were designed and fabricated at IPR to provide required test environments over long durations. A differential temperature measurement based interface detection technique using bulk temperature profiling was also proposed and extensively studied. During these preliminary studies, high reliability and availability was observed for NaK filled remote diaphragm seal type wetted pressure sensor, non-contact pulse radar level sensor and bulk temperature sensor. Performed studies suggest that silicone oil is not suitable as a high-temperature de-coupling fill fluid for long duration applications due to significant thermal expansion. As such, no impact of corrosion was observed on the measurement accuracies of wetted sensors during 1000 hour tests. Performed static tests represent scenarios similar to foreseen measurement applications in LLCS, because all these sensor configurations are planned to be installed in nearly static liquid Pb-Li conditions. Further design optimizations and compatibility of process sensors with environmental factors (like magnetic field, radiation etc.) still need to be addressed for a complete development relevant to applications in heavy liquid metal coolant systems for fusion test blankets. Obtained performance data from validated and optimized sensor configurations maybe useful for RAMI and FMEA analysis at component level for ancillary plant systems related to LLCB TBM applications.

## Acknowledgements

Authors would like to acknowledge the assistance rendered by A. Jaiswal (IPR) and K. Pandya (IPR) for the operation and maintenance of test facilities. In addition, authors deeply appreciate the support from C.S. Sasmal (IPR) for the metallographic work and analysis of samples from wetted sensors.